# An improved scoring matrix for multiple sequence alignment

Jian-Jun SHU*, Kian Yan YONG and Weng Kong CHAN

*School of Mechanical & Aerospace Engineering, Nanyang Technological University, 50 Nanyang Avenue, Singapore 639798*

**ABSTRACT**

The way for performing multiple sequence alignment is based on the criterion of the maximum scored information content computed from a weight matrix, but it is possible to have two or more alignments to have the same highest score leading to ambiguities in selecting the best alignment. This paper addresses this issue by introducing the concept of joint weight matrix to eliminate the randomness in selecting the best multiple sequence alignment. Alignments with equal scores are iteratively rescored with the joint weight matrix of increasing level (nucleotide pairs, triplets and so on) until one single best alignment is eventually found. This method for resolving ambiguity in multiple sequence alignment can be easily implemented by use of the improved scoring matrix.

## 1. Introduction

In the search for DNA regulatory elements such as binding sites, promoter, donor sites, TATA box and genes, the multiple sequences containing these elements have to be aligned against each other. These elements are highly but not absolutely conserved and a weight matrix is used to represent and score the multiple sequences [1]. However, the current motif discovery algorithms based on the weight matrix technique for scoring multiple sequence alignment in terms of information content are not without their limitations [2]. From the analysis of these algorithms, the highest performance coefficient on the binding site level of search is only *30.2%* using Motif Sampler [3], which is an algorithm modified from the widely adopted Gibbs Sampling method [4]. This may be a result of randomness in selecting the best alignment from cases whereby there are multiple peaks. Hence, there are rooms for improvement, which is evident from many different approaches that have been developed [5-9].

In this paper, a method of removing the randomness in selection is proposed. Randomness in selection occurs when there is more than one choice of alignments with the highest information content [10]. If one peak is randomly selected, the accuracy of multiple sequence alignment is compromised. This may be the reason that methods based on applied information theory cannot achieve much higher sensitivity, specificity and performance. For example, by randomly selecting two peaks of similar information content, there is a *50%* chance of selecting the wrong peak.

* Correspondence should be addressed to Jian-Jun SHU, mjjshu@ntu.edu.sg

In order to overcome this problem, a simple method is proposed to eliminate the randomness of peak selection and to provide the best alignment, through the use of joint weight matrix (JWM) in this paper. Its flexibility means that a higher-level JWM can be used to work with cases with multiple peaks. The higher the level of JWM used, the lesser will the number of peaks be, until eventually a single peak is obtained. In this paper, JWM has been shown to reduce successfully the number of peaks in multiple sequence alignment.

## 2. Systems and Methods

The concept of JWM is presented here to demonstrate how two or more ambiguous selections can be reduced. Two sequences are used in this example. The longer one represents the DNA sequence and the shorter one represents a motif sequence, which is aligned to the former. The motif is assumed to be a perfect weight matrix with *100%* base weightage at each position. The score is then either *1* for match or *0* for mismatch at each position for simplicity of demonstration.

Since the sequence is *7* bp (base pair) long and the motif is *4* bp, the total number of possible shift positions without introducing gaps is $7-4+1=4$ in Figure 1. Table 1 shows the sequence alignment.

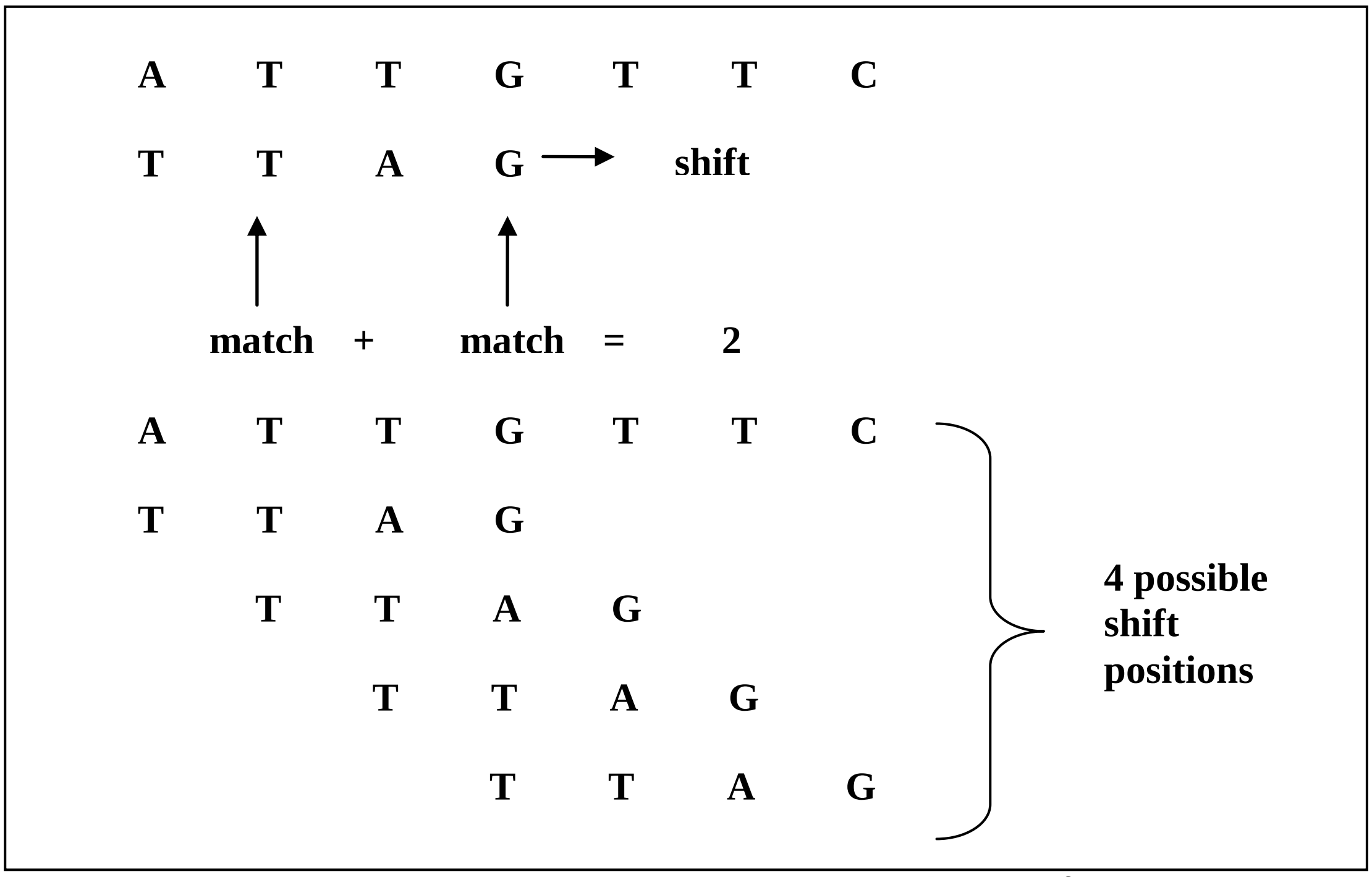

**Figure 1:** Scoring a DNA sequence with a motif

**Table 1:** Tabulated score for match between DNA and motif based on base by base

| Position | 1 | 2 | 3 | 4 | 5 | 6 | 7 |
|---|---|---|---|---|---|---|---|
| DNA | A | T | T | G | T | T | C |
| Motif | T | T | A | G | | | |
| Score | 2 | 2 | 1 | 1 | | | |

The score for the four possible alignments presents an ambiguous choice between positions $1$ and $2$, which are possible alignments with the highest score of $2$. Since there is more than one peak or alignment, the second-level JWM is used to score the alignment. Table 2 shows the result using the second-level comparison.

**Table 2:** Tabulated score of match between DNA and motif based on two bases

| Position | 1 | 2 | 3 | 4 | 5 | 6 | 7 |
|---|---|---|---|---|---|---|---|
| DNA | A | T | T | G | T | T | C |
| Motif | T | T | A | G | | | |
| Score | 0 | 1 | 0 | 0 | | | |

The result clearly shows that between positions $1$ and $2$, the better match for the motif with the DNA is the position $2$ with the matching score of $1$ as compared with the position $1$ with $0$.

## 3. Algorithm

Here it is shown how JWM can be integrated with sequence alignment tool to remove the randomness of selection during the alignment process. The following are the additional steps added using JWM:

**Step 1:** Determine a weight matrix

$$w(b,i) = \frac{n(b,i)}{\sum_{b\in\{\mathrm{A,T,G,C}\}} n(b,i)}, \tag{1}$$

where $n(b,i)$ is the number of each base $b \in \{\mathrm{A,T,G,C}\}$ at each position $i$.

**Step 2:** Calculate the second-level JWM

$$w_2(b_1 b_2, i) = w(b_1, i)\, w(b_2, i+1). \tag{2}$$

For the second-level JWM, the number of possible combinations of the four bases is $4^2 = 16$. Hence JWM is a matrix size of $16$ by window length.

**Step 3:** From the weight matrix, the uncertainty of each combination of bases is

$$Hs(i) = -\sum_{j=1}^{m} \sum_{b_j\in\{\mathrm{A,T,G,C}\}} w_m(b_1 b_2 \cdots b_m, i)\, log_2\, w_m(b_1 b_2 \cdots b_m, i). \tag{3}$$

For the second-level JWM, $m$ is the value of $2$.

**Step 4:** The information content for each base is then

$$R(i) = 2^m - Hs(i) - e[n(i)], \tag{4}$$

where $e[n(i)]$ is a small sample correction for $Hs(i)$ [11].

**Step 5:** The score for one shifting position is then

$$R_{\text{shift}}(\text{sp}) = \sum_i R(i). \tag{5}$$

The shift position (sp) ranges from negative to positive shifting parameter.

**Step 6:** Shift JWM as predetermined to get the alignment score plot of information content versus shifting position. From the alignment score plot, the highest peak is chosen among the ambiguous choice of the previous set of peaks to be generated.

**Step 7:** If there is still ambiguity after using the second-level JWM, a higher-level JWM (three or higher) should be calculated

$$w_m(b_1 b_2 \cdots b_m, i) = w(b_1, i)\ w(b_2, i+1)\ \cdots\ w(b_m, i+m-1). \tag{6}$$

Repeat the Steps 3 to 6 using the higher-level JWM in (6) when there is ambiguity in peak selection if using any lower-level JWM.

## 4. Implementation

An example of how JWM is used to eliminate or reduce ambiguity is shown using data from $16$ randomly generated sequences of $15$ bp (Tables 3 and 4) that bind to OxyR [12]. For illustration purpose, the centre $9$th base is taken to be the start site of transcription, labeled as the position $0$. The alignment score is obtained by using the window of $5$ bases from $-1$ to $+3$ and the range of shifting position set from $-8$ to $+6$ with respect to the start site. The sequences are shifted one base at a time and the new alignment score is recalculated based on the simplified sequence logo [13] in Figure 2(a).

Window and shifting parameters are selected such that an ambiguous choice of more than one peak is resolved. By shifting one of the sequences from $-8$ to $+6$, the alignment score based on window from $-1$ to $+3$ show two peaks at shift positions $-5$ and $0$ in Figure 2(b). From the simplified sequence logo, the information content prior to shifting of any sequence is

$$R_{\text{shift}}(0) = 0.0637 + 0.1950 + 0.5087 + 0.1504 + 0.2500 = 1.1678 \text{ bits}.$$

**Table 3:** Weight matrix of *16* OxyR binding sequences from base positions *-8* to *+6*

| | -8 | -7 | -6 | -5 | -4 | -3 | -2 | **-1** | **0** | **+1** | **+2** | **+3** | +4 | +5 | +6 |
|---|---|---|---|---|---|---|---|---|---|---|---|---|---|---|---|
| 1 | T | C | A | C | A | C | C | **G** | **A** | **C** | **T** | **T** | G | T | G |
| 2 | A | C | T | T | A | T | C | **G** | **A** | **T** | **C** | **C** | G | C | A |
| 3 | C | A | T | T | A | A | C | **A** | **A** | **T** | **A** | **G** | G | G | C |
| 4 | T | A | C | G | A | T | A | **A** | **T** | **A** | **G** | **G** | C | A | A |
| 5 | C | G | T | A | C | A | T | **T** | **A** | **T** | **C** | **C** | A | T | A |
| 6 | C | T | A | T | T | A | T | **T** | **G** | **T** | **A** | **A** | C | A | G |
| 7 | A | C | T | T | T | C | C | **C** | **A** | **G** | **A** | **G** | T | T | C |
| 8 | C | A | G | A | G | A | T | **C** | **G** | **C** | **T** | **C** | T | A | A |
| 9 | A | C | T | A | A | A | C | **T** | **T** | **C** | **T** | **G** | A | T | A |
| 10 | A | G | T | T | A | T | C | **G** | **G** | **T** | **A** | **T** | A | A | T |
| 11 | A | C | G | A | T | G | G | **A** | **A** | **T** | **C** | **C** | A | T | A |
| 12 | C | A | G | A | G | A | T | **C** | **G** | **C** | **T** | **C** | T | A | A |
| 13 | A | T | C | A | C | T | G | **A** | **C** | **T** | **A** | **C** | A | A | T |
| 14 | A | T | T | A | G | C | G | **A** | **T** | **T** | **A** | **C** | C | G | T |
| 15 | A | T | T | A | C | C | T | **A** | **T** | **C** | **G** | **C** | T | G | C |
| 16 | C | T | A | T | T | A | T | **T** | **G** | **T** | **A** | **A** | C | A | G |
| **A** | 8 | 4 | 3 | 8 | 6 | 7 | 1 | **6** | **6** | **1** | **7** | **2** | 5 | 7 | 7 |
| **C** | 6 | 5 | 2 | 1 | 3 | 4 | 6 | **3** | **1** | **5** | **3** | **8** | 4 | 1 | 3 |
| **T** | 2 | 5 | 8 | 6 | 4 | 4 | 6 | **4** | **4** | **9** | **4** | **2** | 4 | 5 | 3 |
| **G** | 0 | 2 | 3 | 1 | 3 | 1 | 3 | **3** | **5** | **1** | **2** | **4** | 3 | 3 | 3 |
| **%A** | 0.50 | 0.25 | 0.19 | 0.50 | 0.38 | 0.44 | 0.06 | **0.38** | **0.38** | **0.06** | **0.44** | **0.13** | 0.31 | 0.44 | 0.44 |
| **%C** | 0.38 | 0.31 | 0.13 | 0.06 | 0.19 | 0.25 | 0.38 | **0.19** | **0.06** | **0.31** | **0.19** | **0.50** | 0.25 | 0.06 | 0.19 |
| **%T** | 0.13 | 0.31 | 0.50 | 0.38 | 0.25 | 0.25 | 0.38 | **0.25** | **0.25** | **0.56** | **0.25** | **0.13** | 0.25 | 0.31 | 0.19 |
| **%G** | 0.00 | 0.13 | 0.19 | 0.06 | 0.19 | 0.06 | 0.19 | **0.19** | **0.31** | **0.06** | **0.13** | **0.25** | 0.19 | 0.19 | 0.19 |

**Table 4:** The second-level JWM for *16* OxyR binding sequences

| | -7 | -5 | -3 | **-1** | **1** | 3 |
|---|---|---|---|---|---|---|
| **AA** | 0.0469 | 0.2109 | 0.0469 | **0.0977** | **0.0273** | 0.0469 |
| **AC** | 0.0469 | 0.1055 | 0.1172 | **0.0391** | **0.0117** | 0.0313 |
| **AT** | 0.1094 | 0.1406 | 0.1406 | **0.0781** | **0.0156** | 0.0313 |
| **AG** | 0.0469 | 0.1055 | 0.0703 | **0.0977** | **0.0078** | 0.0156 |
| **CA** | 0.0586 | 0.0234 | 0.0313 | **0.0781** | **0.1641** | 0.2109 |
| **CC** | 0.0586 | 0.0117 | 0.0781 | **0.0313** | **0.0703** | 0.1406 |
| **CT** | 0.1367 | 0.0156 | 0.0938 | **0.0625** | **0.0938** | 0.1406 |
| **CG** | 0.0586 | 0.0117 | 0.0469 | **0.0781** | **0.0469** | 0.0703 |
| **TA** | 0.0586 | 0.1172 | 0.0391 | **0.0781** | **0.2188** | 0.0469 |
| **TC** | 0.0586 | 0.0586 | 0.0977 | **0.0313** | **0.0938** | 0.0313 |
| **TT** | 0.1367 | 0.0781 | 0.1172 | **0.0625** | **0.1250** | 0.0313 |
| **TG** | 0.0586 | 0.0586 | 0.0586 | **0.0781** | **0.0625** | 0.0156 |
| **GA** | 0.0234 | 0.0234 | 0.0078 | **0.0586** | **0.0273** | 0.0703 |
| **GC** | 0.0234 | 0.0117 | 0.0195 | **0.0234** | **0.0117** | 0.0469 |
| **GT** | 0.0547 | 0.0156 | 0.0234 | **0.0469** | **0.0156** | 0.0469 |
| **GG** | 0.0234 | 0.0117 | 0.0117 | **0.0586** | **0.0078** | 0.0234 |

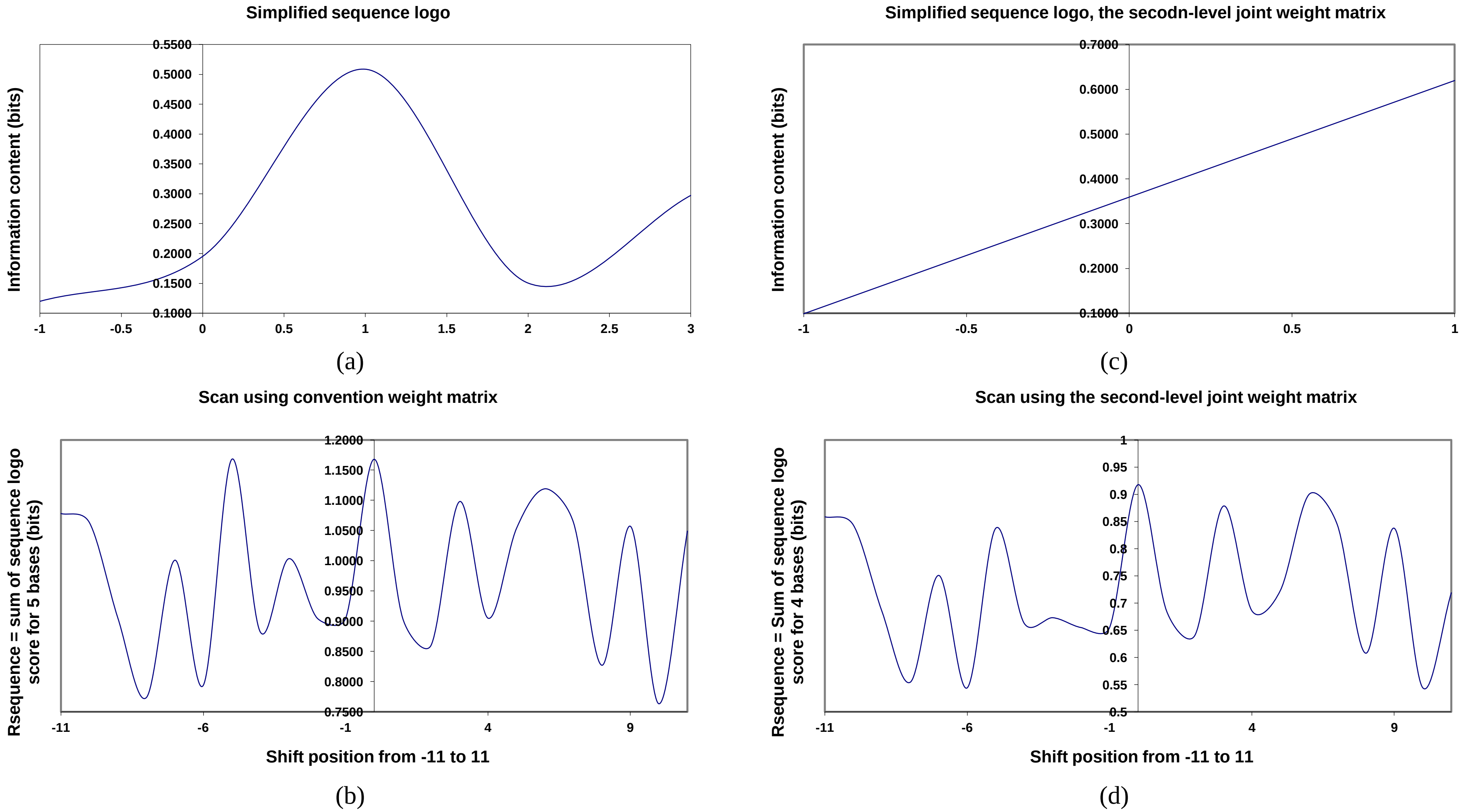

**Figure 2:** Graphical results of OxyR binding sites (a) Simplified sequence logo (b) Information content of each shift positions of a randomly selected sequence using conventional weight matrix (c) Simplified sequence logo (JWM) (d) The same scanning process in (b) using JWM

One of the sequences is randomly selected and shifted about its position. The weight matrix is calculated for each new position and a set of $R_{\text{shift}}(\text{sp})$ is obtained by the end of the shift. The amount of shift required for the $16$ sequences to produce $R_{\text{shift}}$ is plotted in Figure 2(b), where two peaks are located at shift positions $-5$ and $0$. The situation is ambiguous and a higher-level search is required by using JWM. The weight matrix is replaced by the second-level JWM in the new search. The new $R_{\text{shift}}$ plot based on the higher-level JWM is shown in Figure 2(d).

The new alignment score using JWM shows clearly that the shift position $0$ has higher information content, as compared with the shift position $-5$. Hence, the best alignment is the original position $0$. Instead of randomly selecting one of the peaks, it is rational to select the peak with higher information content.

## 5. Discussion and Conclusions

In the selection of the best multiple sequence alignment using the conventional weight matrix, it is assumed that the probability of each base is independent of its neighboring one. Output from a multiple sequence alignment program is not always the same. This can be attributed to several factors. One of the important contributing factors is the conventional scoring matrix. The best alignment at each stage is decided by the highest score with the conventional scoring matrix. However, there are cases whereby there is more than one of such score. This creates an ambiguity in selecting the best alignment. A random choice can be made but it may result in a less than optimal alignment.

The following shows examples of ambiguities found using the conventional scoring matrix. The benchmark database (Table 5) consists of DNA sequences containing amelogenin protein in the study of its origin and evolutionary path [14]. Cases of ambiguity during multiple sequences alignment using the conventional scoring matrix are shown in the Figure 3. For example, the ambiguity is found when the sequence 2 (DMSPARC) at position $0$ and $3$ of window, with window placed at the $18$ th base from the start (first base on the left).

**Table 5:** Ambiguities in sequence alignment using the window size of $5$ bases

| No. | Names | Sequences | Window position | Ambiguity positions |
|---|---|---|---|---|
| 2 | DMSPARC | ATGCGCTCCCTTTGGCTGCTG<br>CTCGGCTTGGGCCTGCTGGC<br>TGTGAGCCACGTCCAGGCCT | 18 | 0,3 |
| 4 | RATSC1 | ATGAAGGCCGTGCTTCTCCT<br>CCTGTATGCCTTGGGGATCG<br>CTGCTGCAGTCCCG | 18 | 0,3 |
| 5 | MOUSESC1 | ATGAAGGCTGTGCTTCTCCTC<br>CTGTGCGCCTTGGGAACCGC<br>TGTGGCAATCCCG | 18 | 0,3 |
| 6 | HUMANHEVIN | ATGAAGACTGGGCCTTTTTTC<br>CTATGTCTCTTGGGAACTGC<br>AGCTGCAATCCCG | 11 | -5,-4 |
| 7 | BOVINAMEX | ATGGGGACCTGGATTTTGTTT | 20 | -4,0 |
| | | GCCTGCCTCCTGGGAGCAGC CTTCTCTATGCCT | | |
| 10 | XENOPUS2AM | ATGAGGCCATTGGTAATGCT AACAGCTCTCATTGGAGCAG CCTTTTCTCTTCCT | 7 | -1,0 |
| 12 | MOUSEAMEX | ATGGGGACCTGGATTTTGTTT GCCTGCCTCCTGGGAGCAGC TTTTGCTATGCCC | 20 | -4,0 |
| 13 | RATAMEX | ATGGGGACCTGGATTTTGTTT GCCTGCCTCCTGGGAGCAGC TTTTGCTATGCCC | 20 | -4,0 |
| 16 | HUMANAMEY | ATGGGGACCTGGATTTTGTTT GCCTGCCTTGTGGGAGCAGC TTTTGCCATGCCT | 20 | -4,0 |
| 17 | HUMAMAMEX | ATGGGGACCTGGATTTTATTT GCCTGCCTCCTGGGAGCAGC TTTTGCCATGCCT | 20 | -4,0 |
| 18 | CAVIAAMEX | ATGGGAACCTGGATTTTGTTT GCCTGCCTCTTGGGAACAGC CTTTGCTATGCCT | 20 | -4,0 |
| 23 | CHICKENSPA | ATGAGAACCTGGATTTTCTTC TTCCTCTGCCTGGCAGGCAA AGCCCTGGCAGCTCCG | 16 | -3,0 |
| 24 | QUAILSPARC | ATGAGAGCCTGGATTTTCTTC CTCCTCTGCCTGGCAGGCAA AGCCCTGGCAGCCCCG | 19 | -3,0 |
| 25 | ZEBRAFSPAR | ATGAGGGTTTGGATCTTCTTC CTGTTCTGCCTCGCTGGCAA GACTCTGGCAGCTCCA | 16 | 0,3 |
| 26 | TROUTSPARC | ATGAGGGTGTGGATTGTCTT CCTCCTGTGCCTAGCTGGTCA GGCATTCACCGCTTCC | 7 | -4,5 |
| 27 | XENOPUSSPA | ATGAGGGTCTGGGTCTTCTTC GTCTTGTGCCTGGCTGGCAA AGCACTAGCTGCCCCT | 16 | 0,3 |

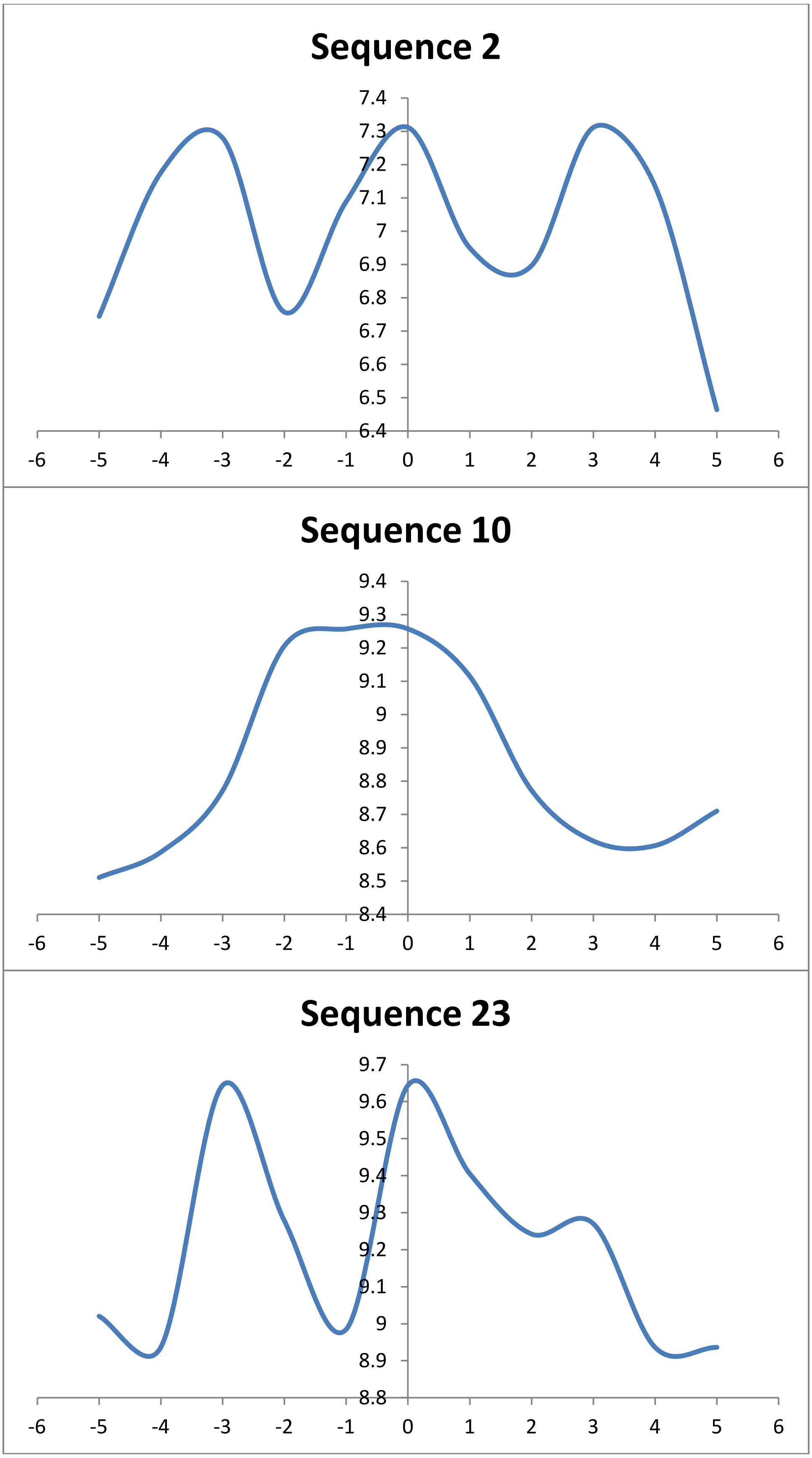

**Figure 3:** Alignment score plot of information content versus shifting position

The examples above shows that ambiguities are frequent enough to be of concern during multiple sequence alignment, which may result in a suboptimal alignment. This problem can be overcome by using the proposed joint weight matrix for scoring. The proposed scoring matrix allows a closer look at each alignment by considering two or more bases for each scoring element. By comparing two bases at one time, the probability of the next base is affected by what appears before it. In fact, there are $16$ probabilities of a pair of bases as compared with just $4$ probabilities if only one base is considered. This increases the depth of search to reduce the number of peaks. Under the Implementation section, it is shown how the second-level JWM can identify the highest peak when a conventional weight matrix could not. This reduces the error that may occur when "conflicts are resolved" by making a "pseudo-random choice" [10].

The higher-level of JWM can be used depending on the level of accuracy required. For example, the second-level JWM may be able to reduce the number of peaks from $5$ to $3$. The randomness is reduced when one is choosing the best peak from $3$ instead of $5$ possible sites. However, if the application requires a level of match to be of greater accuracy, a higher-level of JWM may be needed to proceed. The higher-level of JWM can further filter out more peaks till only one obvious choice is left. Although the higher-level of JWM may require more computation time and additional scan, this may be compensated by the faster convergence of results as a better alignment is selected early in the iterations. This is true especially for cases whereby a large number of iterations are required before a satisfactory convergence can be found [15]. JWM can be used to improve applications using conventional weight matrix system in bioinformatics. Besides aligning DNA sequences, JWM can also be implemented in protein sequence alignment.